\documentclass[aps,pra,twocolumn,superscriptaddress,longbibliography,floatfix]{revtex4-1}
\usepackage{amsfonts}
\usepackage{graphicx}
\usepackage{amsmath}
\usepackage{times}
\usepackage{amssymb}
\usepackage{color}
\usepackage[colorlinks,bookmarks=false,citecolor=blue,linkcolor=red,urlcolor=blue]{hyperref}

\newcommand{\bei}{\begin{itemize}}
\newcommand{\eei}{\end{itemize}}

\begin{document}

\title{
The $S=1/2$ Kagome Heisenberg Antiferromagnet Revisited%
}

\author{Andreas M. L\"auchli}
\affiliation{Institut f\"ur Theoretische Physik, Universit\"at Innsbruck, A-6020 Innsbruck, Austria}
\author{Julien Sudan}
\affiliation{Rue du Stand 16, CH-2053 Cernier, Switzerland}
\author{Roderich Moessner}
\affiliation{Max-Planck-Institut f\"{u}r Physik komplexer Systeme, N\"{o}thnitzer Stra{\ss}e 38, D-01187 Dresden, Germany}

\date{\today}

\begin{abstract}
We examine the perennial quantum 
spin-liquid candidate $S=1/2$ Heisenberg antiferromagnet on the kagome lattice.
Our study is based on achieving  Lanczos diagonalization of the Hamiltonian on a $48$ site cluster in sectors 
with dimensions as a large as $5 \times 10^{11}$.
The results reveal novel intricate structures in the low-lying energy spectrum. 
These structures by no means unambiguously support a $\mathbb{Z}_2$ spin liquid 
ground state, but instead appear compatible with several scenarios, including
four-fold topological degeneracy, inversion symmetry breaking and a combination thereof. 
We discuss finite-size effects, such as the apparent  absence of ETH, and note that while considerably reduced, some are  still present for the  largest cluster.
Finally, we observe that an XXZ model in the Ising limit reproduces remarkably
well the most striking features of finite-size spectra.
\end{abstract}

%\pacs{75.10.Jm, 05.30.Fk, 02.70.-c}
%%02.70.-c		:	Computational techniques; simulations
%%71.10.Fd	:	Lattice fermion models (Hubbard model, etc.)
%%03.67.-a		:	Quantum information
%%67.85.-d,  ultracold Gases
%%05.30.Fk, quantum statistical mechanics
%%75.10.Jm Heisenberg model

\maketitle
%%%%%%%%%%%%%%%%%%%%%%%%%%%%%%%%%%%%%%%
The $S=1/2$ kagome Heisenberg magnet is arguably the best studied, 
yet most enigmatic, candidate spin liquid, identified as such 
in the very early works of the field of highly frustrated 
magnetism~\cite{Elser1989,Zeng1990}. Many different numerical approaches have been applied
to this magnet: exact diagonalisation~\cite{Elser1989,Zeng1990,Chalker1992,Leung1993,Lecheminant1997,Waldtmann1998,Lauchli2011,Nakano2011,Changlani2018,Wietek2019}, 
resonating valence bond physics inspired methods~\cite{Zeng1995,Mila1998,Mambrini2000,Misguich2002,Misguich2003,Poilblanc2010,Schwandt2010,Rousochatzakis2014,Ioannis2018}, 
variational Monte Carlo~\cite{Ran2007,Iqbal2011a,Iqbal2011b,Tay2011,Clark2013,Iqbal2014}, 
coupled cluster treatments~\cite{Goetze2011}, contractor renormalization (CORE)~\cite{Budnik2004,Capponi2004,Capponi2013},
series expansions~\cite{Singh1992,Singh2007}, density matrix renormalization group (DMRG)~\cite{Jiang2008,Yan2011,Depenbrock2012,Jiang2012,Nishimoto2013,He2017}, 
and tensor network algorithms~\cite{Evenbly2010,Xie2014,Mei2017,Jiang2016,Liao2017}.

These have yielded an enigmatic phenomenology, with evidence for the following features: 
(a) all correlations are short-ranged; (b) the singlet gap, if non-vanishing, is numerically tiny; 
(c) there exists a huge number of  (near-)degenerate states apparently not related by symmetry; 
(d) these spectral features are remarkably stable for classes of perturbations around the Heisenberg point. 
On general grounds, it is not clear how to reconcile these, in particular (a) with (b), and (c) with (d). 

Given these confusing signals, it is perhaps not surprising that 
confidence in various pictures of the behaviour of this magnet has ebbed and flowed, 
with new technologies providing invaluable new insights which in turn generate new scenarios
of varying shelflife. An important  breakthrough 
was a DMRG tour de force by Yan and coworkers~\cite{Yan2011}
and a number of follow-on studies using that method~\cite{Depenbrock2012,Jiang2012},
as it turns into a tool for the study of two-dimensional magnets.
Taken together, these suggested as the most likely scenario of a $\mathbb{Z}_2$ gapped spin liquid, 
based on the following  
observations: correlations appear to be short-ranged, with candidate ordering
patterns imposed at the boundaries decaying swiftly into the bulk. Evaluations of the universal
contribution to the entanglement
entropy show, with differing degrees of confidence, the value expected for this topological state~\cite{Jiang2012,Depenbrock2012}. 

A source of uncertainty hard to quantify in this evidence lies in the fact that  
DMRG is not  an unbiased method, preferring
low-entanglement states over highly entangled ones, so that 
the last word may very well not have been spoken. Indeed, a more recent DMRG study employing flux threading found evidence for a much smaller spin 
gap than that given by previous DMRG estimates, and suggested the possibility of a U(1) spin liquid with an excitation spectrum containing (gapless) Dirac cones \cite{He2017}. 

Against this background, the work reported here revisits possible alternative scenarios. The new material
presented here is based on state-of-the-art exact diagonalisation work. This is in the tradition of
exact diagonalisation studies that have historically been a lynchpin of the study of kagome; their main advantage is that
they are numerically exact and unbiased, while providing a comprehensive picture of the low-energy physics, including
the quantum numbers of the excitation spectrum above the ground state. The main disadvantage lies in the limitation 
to finite sizes, which even given Moore's law, is only being pushed back slowly. 

The quantifiable technical advance lies in our capacity to treat a cluster of 
48 sites with a Hilbert space dimension of $2^{48}\approx 2.8 \times 10^{14}$.  
Using a highly optimized, message-passing based exact 
diagonalization (ED) code it has been possible to obtain   the low-lying
energy spectrum in symmetry sectors comprising up to $\approx 5\times 10^{11}$ states. 
To the best of our knowledge, this is among the largest 
number of $S=1/2$ spins treated in exact diagonalization in a comparable context.
%, and this also
%raises the bar for the demonstration of `supremacy' of 
%quantum simulations over classical computers.

%%%%%%%%%%%%%%%%%%%%%%%%%%%%%%%%%%%%%%%
\begin{figure*}[t]
\begin{center}
\includegraphics[width=\linewidth]{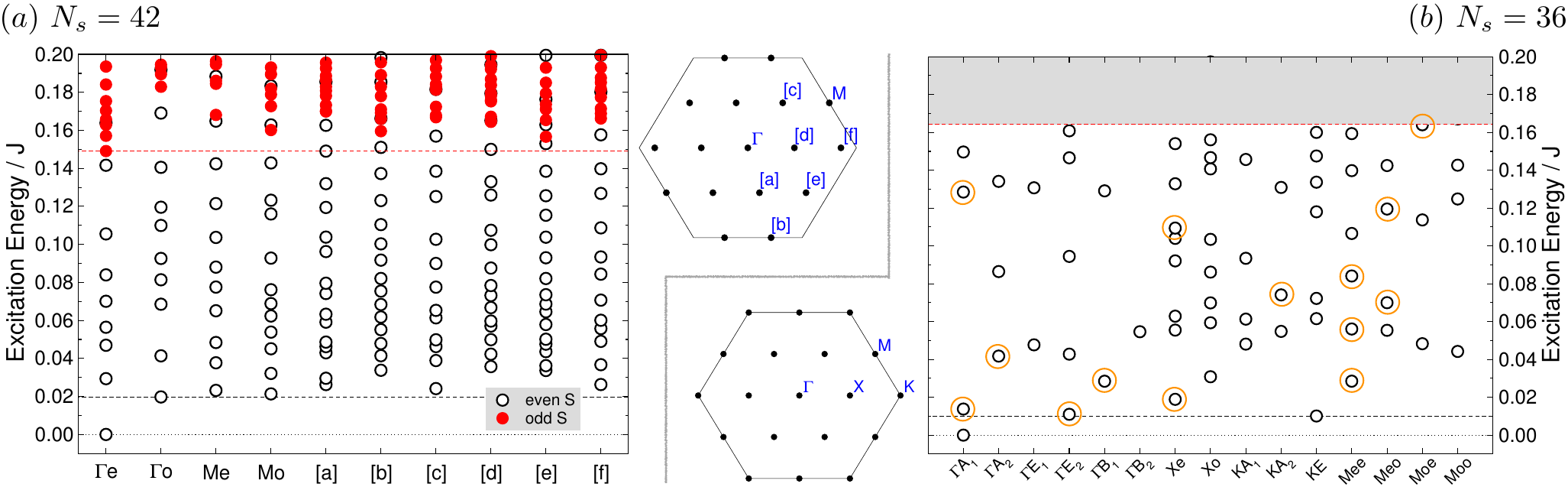}
\caption{(Color online)
Low-energy Lanczos spectra of the $S=1/2$ Heisenberg model for clusters with (a) $N_s=42$  and  (b) $N_s=36$ sites.
The black, empty symbols denote singlet levels, while the the red, full symbols indicate triplet levels. The black and red dashed lines indicate the location of the singlet-singlet and singlet-triplet gap. The
spectrum for $N_s=42$ is only totally converged for the lowest level in each symmetry sector. We nevertheless plot the complete spectrum of the tridiagonal
Lanczos matrix in order to provide a visual impression of the build-up of a comparatively high density of states in the low-energy spectrum. The spectrum 
shown for $N_s=36$ is fully converged. In the center we display the Brillouin zones of the two clusters including the labelling of the discrete k-points. The additional orange circles in (b) denote eigenstates with pronounced dimer-dimer correlations, see main text for details.
}
\label{fig:lowergspectrum42_36}
\end{center}
\end{figure*}
%%%%%%%%%%%%%%%%%%%%%%%%%%%%%%%%%%%%%%%

Physically, the 48 site cluster  has the following important properties. First, it
is a highly symmetric cluster. Second,  it is compatible with many of the principal proposed ordering patterns. And, third,
it severely reduces finite-size effects by eliminating a large class of winding loops (loops on the lattice
winding around the periodic boundary) of length $L=8$ present for the hitherto largest-studied 36- and 42-site clusters~\footnote{
The $36$, $42$ and $48$ site clusters have respectively $54$, $35$ and $12$ excess length-8 loops, i.e.~loops which have a 
nontrivial winding number around the torus.}.

In the following, we first report the new data on energies and gaps, which
can act as benchmark and reference for the future. For the 48 site cluster, the ED ground state energy is comfortably 
below that determined from  DMRG. Next, we  discuss the 
structure of the low-lying energy spectrum, which turns out to be 
 consistent with inversion symmetry breaking, or with the presence of a topological degeneracy.
 We provide a detailed analysis of correlations for large system, 
 both spin-spin and energy-energy (dimer-dimer) correlators. We discuss different finite-size 
 effects, most importantly  an apparent absence of eigenstate thermalisation, and structural shifts of levels with 
 respect to each other.
Finally, we identify the XXZ antiferromagnet in the Ising limit, $J_{xy}=1, \Delta\rightarrow \infty$ (or equivalently $j\equiv J_{xy}/J_z\ll1$),
with its considerably reduced Hilbert space, as an effective model for the low-energy sector of the Heisenberg magnet, thereby extending the stability of 
its behaviour to the full range of quantum models $0<j\leq\infty$~\cite{Lauchli2015,YCHe2015}. We conclude with a discussion.

\section{Model and Method}

We investigate the $S=1/2$ antiferromagnetic Heisenberg model
on the kagome lattice
\begin{equation}
H= J \sum_{\langle i,j\rangle} \mathbf{S}_i \cdot \mathbf{S}_j\ ,
\label{eqn:KagomeHamiltonian}
\end{equation}
with the coupling constant set to $J=1$. We investigate the low-lying energy spectrum of finite kagome 
clusters with periodic boundary conditions with $N_s=36$, $42$ and $48$ sites~\footnote{We chose the
$36d$ and $42b$ clusters in the notation of Ref.~\cite{Lauchli2011}, while the 48 site cluster corresponds
to a highly symmetric $4\times4$ unit cells torus.}. We apply a newly developed massively
parallel exact diagonalization code to study these systems, tackling Hilbert spaces of up to $5\times 10^{11}$ basis 
states. Convergence within a few hundred iterations is typically reached for the lowest two to three eigenstates in 
each symmetry sector.

Energy spectroscopy is a  powerful technique to diagnose various states of quantum matter. A characteristic 
"Tower of States" accompanies 
continuous and discrete symmetry breaking, the effective theory describing 
quantum critical points in 1D and 2D~\cite{Schuler2016} can be accessed this way, and ground 
state degeneracy of  topological origin are also directly visible. 

Given the earlier DMRG evidence 
for $\mathbb{Z}_2$ topological order in the kagome AFM, it thus appears highly desirable to
evidence the required fourfold ground state degeneracy in ED as well. 
In the recent activity on chiral spin liquids~\cite{Bauer2014,He2014,Gong2014,Wietek2015} it
has been possible to observe the two- (or four-) fold ground state degeneracy even with modest system sizes 
accessible by ED, while DMRG simulations for the kagome Heisenberg antiferromagnet have failed to report the required
ground state degeneracy so far.  
In one of the simplest RVB states on the kagome lattice originating from the quantum dimer model of 
Misguich {\it et al.}~\cite{Misguich2002}, we would expect two lying levels each at the $\Gamma$ and the (unique) $M$ point for $N_s=42$, 
while four levels at the $\Gamma$ 
point are expected for $N_s=36$ and $N_s=48$~\footnote{The precise sectors in the QDM are as follows:
$N_s=36,48$: $3\times \Gamma_{ee}, 1\times \Gamma_{oe}$ (expressed as $D_{6}$ representations: $2\times \Gamma_{A1}, 1\times \Gamma_{E2}$)
$N_s=42$b: $2\times \Gamma_{e}, 2\times M_{o}$.}

\section{$N_s=42$ site spectrum} Let us first discuss the symmetry sector resolved low-energy 
spectrum of the $N_s=42$ site cluster. The ground state energy and spin gap of this cluster
have been reported previously~\cite{Lauchli2011,Nakano2011}, but not the momentum and lattice $\pi$-rotation
resolved low-energy spectrum. We display the spectrum in  panel (a) of Fig.~\ref{fig:lowergspectrum42_36}.
For comparison we show the classic~\cite{Waldtmann1998} low-energy spectrum of the highly symmetric 
$N_s=36$ site sample in panel (b) on the same scale. 

The kagome antiferromagnet is notorious for its rather dense low-energy spectrum~\cite{Lecheminant1997,Waldtmann1998,Mila1998}. 
In the $N_s=36$ sample there is a subspace (including degeneracy) of about two hundred singlets below the first triplet 
in the spectrum, with the singlet~singlet gap about $0.01011J$. Despite not being able to fully converge all the singlets before the first
triplet in the $N_s=42$ case, it is nevertheless apparent that the low-energy spectrum is still very dense. Taking the number of all
the approximate eigenvalues below the triplet gap as a lower bound for the exact number of singlets, we obtain at least 160 states.
The singlet-singlet gap is $0.01974J$, which remarkably is almost two times larger than for $N_s=36$.
Furthermore no obvious separation of a low-lying set of multiplets forming the ground space and the rest of the spectrum is visible. 

\section{$N_s=48$ site spectrum}
%%%%%%%%%%%%%%%%%%%%%%%%%%%%%%%%%%%%%%%
\begin{figure}
\begin{center}
\includegraphics[width=\linewidth]{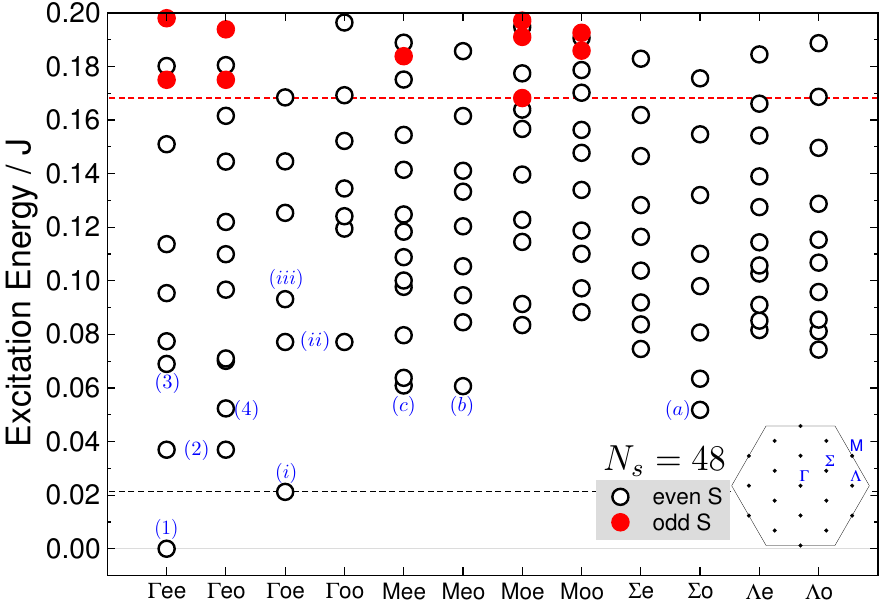}
\caption{(Color online)
Low-energy Lanczos spectrum of the $S=1/2$ Heisenberg model for a cluster with $N_s=48$ sites.
Black and red dashed lines indicate the location of the singlet-singlet and singlet-triplet gap. The
spectrum for is only completely converged for the lowest level in each symmetry sector. We nevertheless plot the complete spectrum of the final 
tridiagonal Lanczos matrix in order to provide an approximate visual impression of the low-energy spectrum. Labeled energy levels  
are discussed in the main text.}
\label{fig:lowergspectrum48}
\end{center}
\end{figure}
%%%%%%%%%%%%%%%%%%%%%%%%%%%%%%%%%%%%%%%
To break down the full Hilbert space into manageable sectors,
we use total $S^z$ conservation, the spin flip symmetry, translation and point group symmetries in order, obtaining
 tractable subspaces of dimension up to $\approx 5\times 10^{11}$. It is currently not possible to simultaneously
exploit the complete SU(2) symmetry group {\em and} the lattice space group in large scale exact diagonalizations.
For performance reasons we only use a subset of the full $D_6$ point group, generated by the $\pi$-rotation around the 
centre of a hexagon, as well as either a reflection along the $x$ or $y$ axis, depending on the momentum sector under 
consideration. When labelling spatial symmetry sectors we state the momentum sector, followed by the eigenvalue $+1$(e) or $-1$(o)
of the $\pi$-rotation and/or the reflection. Using this smaller symmetry group it is nevertheless possible to identify the representation of $D_6$ by a
compatibility table of the representations of the two symmetry groups. We have been able to obtain the lowest energy in all $S^z=0$ sectors with
even spin parity and all but four in the odd spin parity sectors. The available energies are listed in table~\ref{tab:kagome48energies} in the appendix. 
The ground state energy of the $48$-site cluster is $E/NJ=-0.438\ 703\ 897\ 156$, almost half a percent
lower than some of the previous DMRG studies for the same 
cluster~\cite{Jiang2008,Depenbrock2012}. Based on the available triplet sectors our estimate for the spin gap is $\Delta_{S=1}/J=0.168\ 217$, while
the singlet-singlet gap is $\Delta_{S=0}/J=0.021\ 217$.  The spin gap is comparable to earlier CORE~\cite{Capponi2004} and variational results~\cite{Iqbal2014}
on the same system size~\footnote{In the variational calculation~\cite{Iqbal2014} the $S=2$ gap was determined. Our $S=1$ gap is about half as large as their
$S=2$ gap}. 

The low-energy spectrum of the $48$-site cluster is shown in Fig.~\ref{fig:lowergspectrum48}.
In an ideal $\mathbb{Z}_2$ spin liquid situation with a short correlation length one would expect an approximate, but clear-cut, 4-fold ground state 
degeneracy, with a gap to all further excitations. This is not what we observe here, implying that the spin liquid state of the kagome Heisenberg antiferromagnet is
either a $\mathbb{Z}_2$ spin liquid, but with significantly larger correlation lengths than anticipated based on the previous DMRG studies, or 
we are observing a more complex spin liquid state. While we are not able to pinpoint which scenario is realized based on the available system sizes, 
there
nevertheless are a few pointers for the largest system size. In Fig.~\ref{fig:lowergspectrum48} we have labelled 
the 4 expected energy levels for one of the $\mathbb{Z}_2$ spin liquids with the labels $(1),(2)$ [an exact doublet] and $(3)$, all of them at $\Gamma$ point in the 
Brillouin zone. Curiously the first excited state is not part of this set of levels, but seems to be part of an energy-shifted "shadow" structure of levels $(i)$ to $(iii)$ which  differs
from $(1)$ to $(3)$ in their odd quantum number with respect to $\pi$ lattice rotations. It is also worth pointing out that the lowest singlet excitations at {\em finite}
 momentum [e.g. levels (a)-(c)] are at comparatively high energies of $\approx 0.05J$ and above. This is in stark contrast to the $36$- and $42$-site samples, where the lowest
 finite momentum levels are either the first excited overall, or very close in energy,  Fig.~\ref{fig:lowergspectrum42_36}.
%%%%%%%%%%%%%%%%%%%%%%%%%%%%%%%%%%%%%%%
\begin{figure*}
\begin{center}
\includegraphics[width=0.7\linewidth]{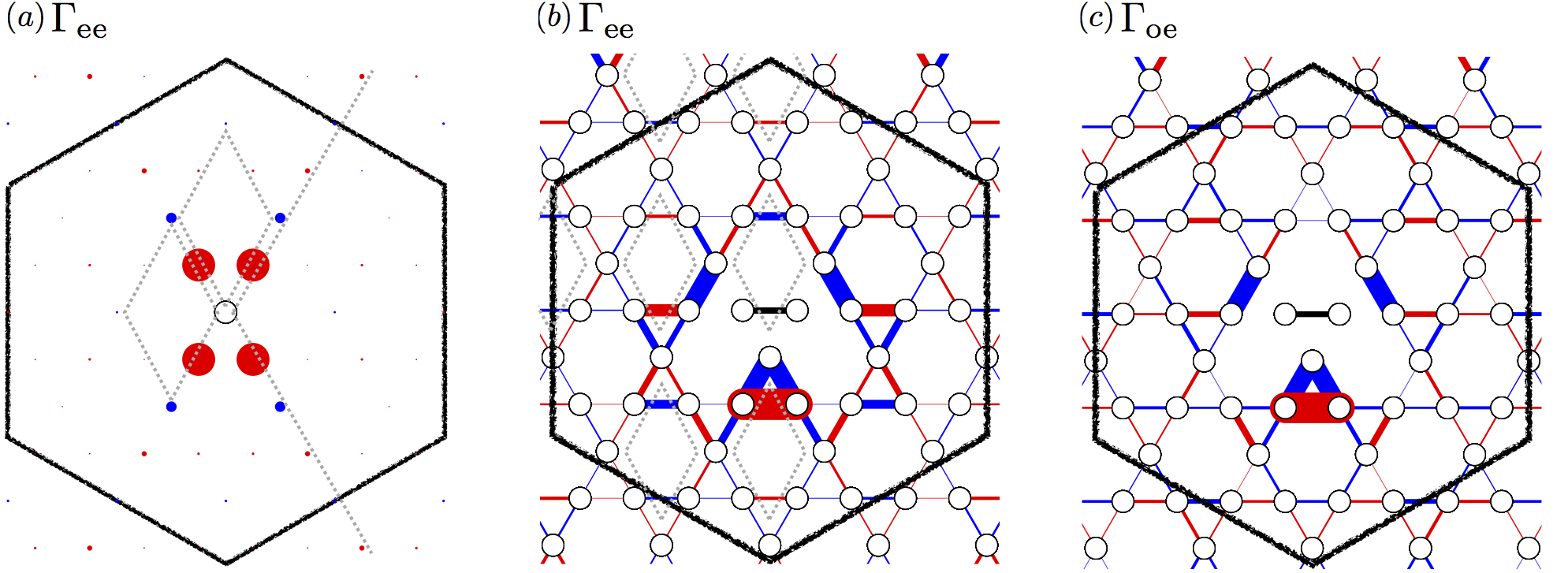}
\end{center}
\caption{(Color online)
Selected correlators for the $48$-site cluster. Filled red (blue) objects denote negative (positive) correlations. Diameters
are proportional to correlation strength.
(a) $\langle S^z_0 S^z_{i}\rangle$ correlations in the ground state. The empty circle denotes the reference site. 
(b) connected $\langle(S^z_i S^z_j)(S^z_k S^z_l)\rangle-\langle (S^z_i S^z_{j})\rangle\langle (S^z_k S^z_{l})\rangle$ 
nearest-neighbor "dimer" correlations in the ground state. The black bond denotes the reference bond.
(c) connected "dimer" correlations in the first excited state ($\Gamma_\mathrm{oe}$ sector).
}
\label{fig:corrs48}
\end{figure*}
%%%%%%%%%%%%%%%%%%%%%%%%%%%%%%%%%%%%%%%

\section{$N_s=48$ site correlation functions}
In order to explore whether the lowest excited state -- located in the $\Gamma_\mathrm{oe}$ sector -- is related to 
a rotation symmetry breaking tendency, we have calculated selected correlations functions in the ground state and the first
level in the $\Gamma_\mathrm{oe}$ sector. Fig.~\ref{fig:corrs48}(a)  displays $\langle S^z_0 S^z_{i}\rangle$ 
in the ground state. As in previous work~\cite{Lauchli2011}, we  find that the strongest spin-spin correlations are not around the
hexagon to which the reference site belongs, but instead along the path which connects the reference
site with its image under periodic boundary conditions (indicated by the straight dashed line). Another interesting structure is the (weak)
staggered correlation signal along a {\em diamond} path (indicated by a dashed diamond lozenge). The correlations in the first excited
state are not shown, but are also weak apart from the wrapping path. In Fig.~\ref{fig:corrs48}(b) and (c) we display the connected "$S^zS^z$-dimer"
correlations: $\mathcal{C}^{zzzz}(i,j,k,l)=\langle(S^z_i S^z_j)(S^z_k S^z_l)\rangle-\langle (S^z_i S^z_{j})\rangle\langle (S^z_k S^z_{l})\rangle$, where
$(i,j)$ and $(k,l)$ denote nearest neighbor bonds. These are diagonal in the computational basis and therefore
computationally friendlier for the large Hilbert spaces under consideration. In the ground state [panel (b)] the correlations show some
interesting structure at short and intermediate distances. We observe a correlation sign pattern which is largely in agreement with a diamond 
valence bond crystal, first discussed in the DMRG study~\cite{Yan2011}, and more recently found to be a stable phase in an extended Heisenberg
model including ferromagnetic further neighbor couplings~\cite{Wietek2019}. The first excited state in the sector $\Gamma_\mathrm{oe}$
[panel (c)]  also exhibits sizeable correlations, with the signs of many correlators changed compared to the ground state. We thus
do not find evidence for a  valence bond type symmetry breaking tendency.

%%%%%%%%%%%%%%%%%%%%%%%%%%%%%%%%%%%%%%%
\begin{figure}[b]
\begin{center}
\includegraphics[width=0.93\linewidth]{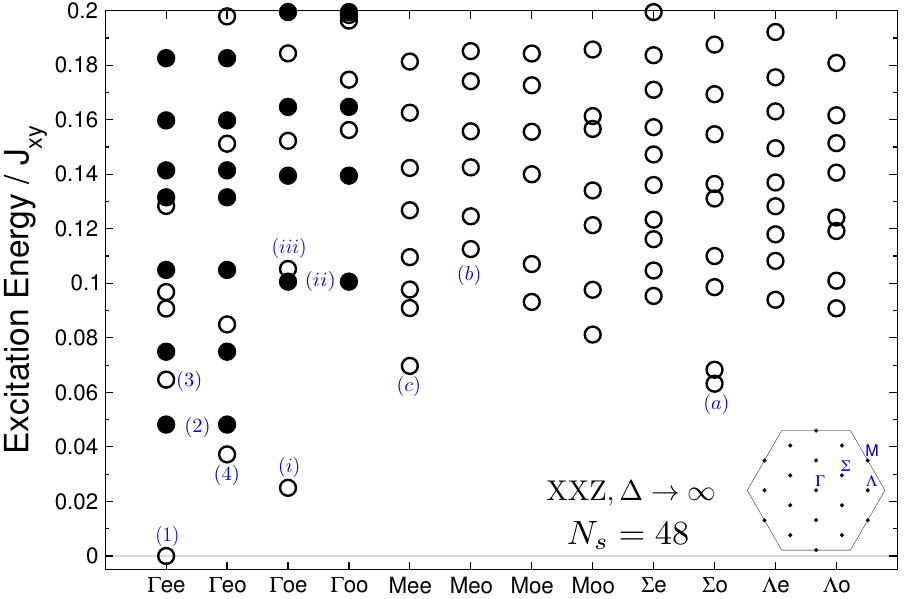}
\caption{(Color online)
XY perturbed AF Ising Model (large $\Delta$ XXZ limit) of the $N_s=48$ site kagome system. 
}
\label{fig:lowergspectrum_XXZ_effective}
\end{center}
\end{figure}
%%%%%%%%%%%%%%%%%%%%%%%%%%%%%%%%%%%%%%%

\section{Low-lying singlet levels}
The large number of low-lying singlets is a hallmark feature in ED studies of the kagome Heisenberg antiferromagnet.
Despite the long history of the problem, the nature of the singlets and a {\em quantitive} effective Hamiltonian describing their energetics 
have been elusive. Here we provide a fresh perspective on these questions. First we have determined 
the nearest-neighbor dimer-dimer correlations in all the singlet eigenstates of the $N_s=36$ below the spin gap. In Fig.~\ref{fig:lowergspectrum42_36}(b)
we highlight those levels with an orange circle which exhibit particularly strong dimer-dimer correlations (presented in detail in
Fig.~\ref{fig:dimercorrs36} in the appendix). The fact that these are broadly scattered across the investigated
energy range is a strong indication that the eigenstate thermalization hypothesis (ETH)~\cite{Srednicki1994} is not (yet) operative. While this is not unexpected at the boundaries
of a many-body spectrum, it is puzzling nevertheless, since the level spacing is already quite small, reminiscent of the situation in the inner part of a many-body
energy spectrum. 

\section{Effective Hamiltonian}
In a recent work two of us have uncovered a striking stability of the energy spectrum of the
Heisenberg antiferromagnet as one moves towards the easy-plane XY limit~\cite{Lauchli2015}.
Furthermore there is also a remarkable continuity towards the Ising limit perturbed with in-plane
exchange, as observed in ground state properties of DMRG simulations~\cite{YCHe2015}. 
This limit has the interesting property that the effective Hilbert space is reduced, because only 
the AF Ising ground states of the kagome lattice need to be retained. For the 48 site cluster this
amounts to a reduction by a factor $\sim 1000$ in total Hilbert space size.
In Fig.~\ref{fig:lowergspectrum_XXZ_effective} we present the energy spectrum of the
XY-exchange perturbed AF Ising model to first order in degenerate perturbation theory for the 48 site
cluster. Interestingly many features of the Heisenberg singlet spectrum of Fig.~\ref{fig:lowergspectrum48} 
can be found here as well. For example the approximate multiplets (1) to (3) and (i) to (iii) are found at similar
locations in the spectrum. Furthermore the lowest finite-momentum excitations (a) and (c) are also low in energy
in the effective model. However there are also some differences, for example the level (4) [(b)] is pushed down [up] 
somewhat when going from the Heisenberg spectrum to the effective XXZ model. Overall we feel that the XY-perturbed
AF Ising configurations on the kagome lattice yields a useful effective Hamiltonian, which is actually able to reproduce 
many features of the low-energy spectrum of the Heisenberg antiferromagnet, and which might be pushed to larger system
sizes, thereby possibly revealing the true nature of the ground state of the  kagome Heisenberg antiferromagnet.

\section{Conclusions} Even for the highly symmetric large $N_s=48$ site cluster,  
no clear $\mathbb{Z}_2$ spin liquid evidence emerges. Also, correlations and spectra at finite wavevector, 
do not suggest  valence-bond ordering. Absent a quantitative understanding of how a U(1) spin liquid and its Dirac cones would show up
in the finite-size spectrum on a torus, we cannot judge the likely validity of this new scenario \cite{He2017}.
While finite-size effects are clearly still present in our results,
some features nonetheless 
demand special attention. For instance, one might have expected the gross features of the physics 
in the ground state to prevail among the lowest excited states, as it is the case 
for ordered magnets or valence bond crystals. 
The absence of ETH in the dimer-dimer correlations then rather suggests that the low-lying singlets of the kagome antiferromagnet
are not just a "soup" of featureless singlets, but instead seem to host a large number of possibly competing many-body states. 
In such a setting, even mildly suboptimal energies obtained variationally 
 may reflect correlations in the trial state a long way from those of the true ground state -- e.g., an error of only 0.5\% in the ground state 
 energy of a 48 site cluster amounts to several times its singlet-singlet gap, a region which hosts quite a number of many-body levels. 
These
aspects clearly require further study. With ongoing progress on several fronts -- numerically (not least DMRG and ED), field-theoretically, 
and with effective models --  the emergence of a consistent picture may perhaps not prove to be quite so elusive in the foreseeable future. 

\begin{acknowledgments}
We are grateful to Yin-chen He, Frank Pollmann and Alexander Wietek for various discussions, and acknowledge
the generous support by the Max Planck Computing Centre in Garching. 
The simulations were performed on the BlueGene/P and on the PKS-AIMS cluster at the MPG RZ Garching, as well
as on the MACH SGI Altix UV machine operated by Uni Innsbruck  and Uni Linz. AML  acknowledges support by the 
Austrian Science Fund (FWF) through DFG-FOR1807 (I-2868) and the SFB FoQus (F-4018). RM acknowledges
DFG support via SFB 1143.
\end{acknowledgments}

\appendix
\section{Energy Spectrum for $N_s=48$}
In table~\ref{tab:kagome48energies} we list the lowest energy in each of the
targeted sectors for future reference. 
\begin{table}[h]
\begin{tabular}{|c|c|c|c|c|}
\hline
momentum & $R_\pi$ & $\sigma_x$ ($\sigma_y$) & $S_\mathrm{total}$ & $E/J$\\
\hline
\hline
$\Gamma$ & $+1$ & $+1$ & even & $-$21.057 787 063\\									     
$\Gamma$ & $+1$ & $-1$ & even & $-$21.020 818 760 \\							   
$\Gamma$ & $-1$ & $+1$ & even & $-$21.036 569 782 \\
$\Gamma$ & $-1$ & $-1$ & even & $-$20.980 603 362 \\
$M$ & $+1$ & $+1$ & even & $-$20.996 851 415 \\
$M$ & $+1$ & $-1$ & even & $-$20.997 096 022  \\
$M$ & $-1$ & $+1$ & even & $-$20.974 317 519  \\
$M$ & $-1$ & $-1$ & even & $-$20.969 472 027 \\
$\Sigma$ & $\times$ & $+1$ & even &$-$20.983 214 \phantom{000}\\
$\Sigma$& $\times$ & $-1$ & even & $-$21.005 970 \phantom{000}\\
$\Lambda$ & $\times$ & $+1$ & even & $-$20.976 185 \phantom{000}\\
$\Lambda$ & $\times$ & $-1$ & even & $-$20.983 468 \phantom{000}\\
\hline
\hline
$\Gamma$ & $+1$ & $+1$ & odd & $-$20.882 732 807\\
$\Gamma$ & $+1$ & $-1$ & odd &  $-$20.882 732 807\\
$\Gamma$ & $-1$ & $+1$ & odd &  $-$20.856 149 771\\	
$\Gamma$ & $-1$ & $-1$  & odd &  $-$20.854 596 491\\
$M$ & $+1$ & $+1$ & odd & $-$20.873 944 088\\
$M$ & $+1$ & $-1$ & odd & $-$20.848 993 609\\
$M$ & $-1$ & $+1$ & odd & $-$20.889 569 935 \\
$M$ & $-1$ & $-1$ & odd & $-$20.871 871 569 \\
$\Sigma$ & $\times$ & $+1$ & odd &N/A\\
$\Sigma$ & $\times$ & $-1$ & odd & N/A\\
$\Lambda$ & $\times$ & $+1$ & odd & N/A\\
$\Lambda$ & $\times$ & $-1$ & odd & N/A\\

\hline
\end{tabular}
\caption{
Lowest energy in each spatial and spin rotation symmetry sector considered for
the $N_s=48$ site kagome Heisenberg cluster.
}
\label{tab:kagome48energies}
\end{table}

\section{Dimer-Dimer correlation functions in selected eigenstates for $N_s=36$}
\label{app:dimerdimer36}
%%%%%%%%%%%%%%%%%%%%%%%%%%%%%%%%%%%%%%%
\begin{figure*}[h]
\begin{center}
\includegraphics[width=0.8\linewidth]{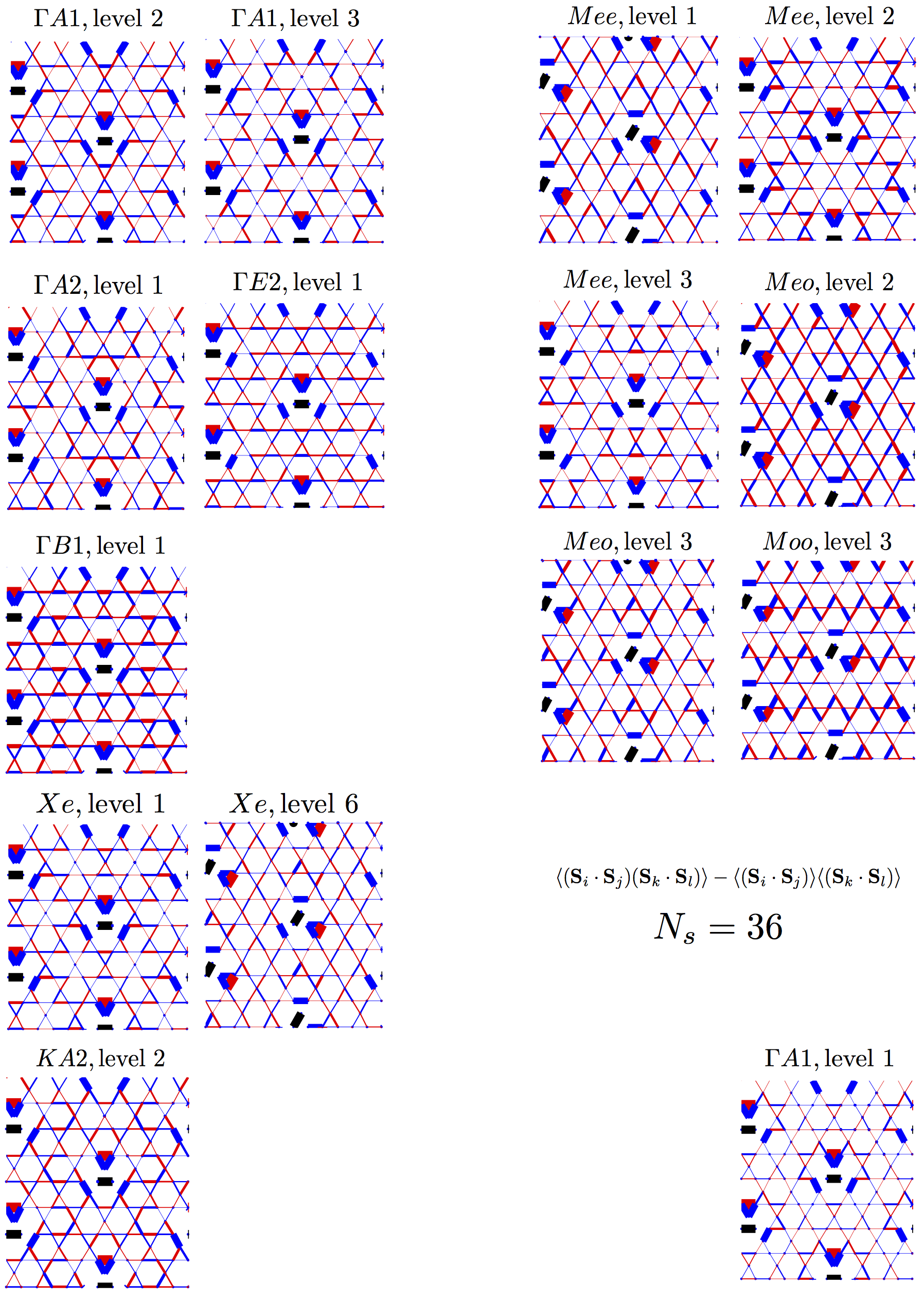}
\end{center}
\caption{(Color online) 
Panel of $N_s=36$ dimer-dimer singlet correlations (defined as indicated in the figure) in the states highlighted by orange circles in Fig~\ref{fig:lowergspectrum42_36}(b).
For comparison we show the ground state dimer-dimer correlations in the bottom right subplot.
}
\label{fig:dimercorrs36}
\end{figure*}
%%%%%%%%%%%%%%%%%%%%%%%%%%%%%%%%%%%%%%%

\bibliography{kagome,biblio}

\end{document}